\documentclass[11pt]{article}

\usepackage{amsmath}
\usepackage{graphicx}
\usepackage{indentfirst}
\usepackage{amssymb}
\usepackage{cite}
\usepackage{color}
\usepackage{subfigure}
\usepackage{varwidth}

\begin{document}

\title{Polarization Images of Solitonic Boson Stars}

\date{}
\maketitle

\begin{center}
\author{Xiao-Xiong Zeng,}$^{a}$\footnote{E-mail: xxzengphysics@163.com}
\author{Chen-Yu Yang,}$^{b}$\footnote{E-mail: chenyuyang2024@163.com}
\author{Hao Yu,}$^{c}$\footnote{E-mail:  yuhaocd@cqu.edu.cn}
\author{Ke-Jian He,}$^{b}$\footnote{E-mail: kjhe94@.163com(Corresponding author)}
\\

\vskip 0.25in
$^{a}$\it{College of Physics and Electronic Engineering, Chongqing Normal University, Chongqing 401331, China}\\
$^{b}$\it{Department of Mechanics, Chongqing Jiaotong University, Chongqing 400074, China}\\
$^{c}$\it{Physics Department, Chongqing University, Chongqing 401331, China}\\
\end{center}

\vskip 0.6in
{\abstract
{ This study investigates the polarization characteristics of solitonic boson stars surrounded by a thin accretion disk. By comparing their polarization images with corresponding optical images, we find a positive correlation between the polarization intensity distribution in the polarization images and the brightness in the optical images. Consequently, the strongest polarization occurs at the location corresponding to the direct image. The influence of the coupling strength of the sixtic potential on the polarization intensity distribution is not monotonic, under strong coupling, the polarization will concentrated on the left side of the image as the coupling strength increases, whereas under weak coupling, it is more evenly distributed across the entire direct image as the coupling strength increases. Moreover, we find that as the initial scalar field increases, both the lensing image and photon ring become more prominent. However, the polarization intensity at these regions remains weak. Due to the absence of the event horizon in solitonic boson stars, the polarization vector can penetrate the stellar interior, unlike in black holes, where no polarization signals exist within the event horizon. Our numerical simulations clearly reveal this phenomenon, suggesting that polarization features may serve as an effective tool for distinguishing solitonic boson stars from black holes.}}
\thispagestyle{empty}
\newpage
\setcounter{page}{1}\

\section{Introduction}
\label{sec:intro}
In recent years, astronomical observations have provided compelling evidence for the existence of ultra-compact objects (stars) in the universe. These observations include the gravitational wave signals from binary mergers detected by the LIGO-Virgo-KAGRA (LVK) collaboration \cite{LIGOScientific:2016aoc,KAGRA:2021vkt}, the 230 GHz imaging observations of the M87* black hole and the Sgr A* black hole by the Event Horizon Telescope (EHT) collaboration \cite{EventHorizonTelescope:2019dse,EventHorizonTelescope:2022wkp,EventHorizonTelescope:2021bee}, and the observations of infrared flares near the Galactic Center by the GRAVITY collaboration \cite{GRAVITY:2020lpa,GRAVITY:2023avo}. They are highly consistent with the theoretical predictions of ultra-compact objects, particularly black holes, within the framework of general relativity.

Although various black hole models can successfully explain these observational phenomena, physicists are not entirely satisfied with these results either because the intrinsic singularity of black holes~\cite{Penrose:1969pc} introduces certain incompleteness from both mathematical and physical perspectives~\cite{Romero:2012ag}. Furthermore, the presence of the event horizon leads to information non-conservation, preventing quantum gravity theories from maintaining unitarity~\cite{Hawking:1976ra}. To avoid these drawbacks of black holes and explain observations in astronomy, the concept of exotic compact objects (ECOs) was introduced~\cite{Cardoso:2019rvt}. Some ECO models exhibit physical characteristics remarkably similar to those of black holes and thus are referred to as mimickers.

Among these ECO models, boson stars have attracted considerable attention. The Kerr hypothesis states that, under appropriate astrophysical conditions, the end-state of complete gravitational collapse forms a spinning and electrically-neutral black hole~\cite{Kerr:1963ud,Penrose:1964wq,Yagi:2016jml,Will:2014kxa}. However, with the self-interaction potential, boson stars can form bound states even in the absence of strong gravity~\cite{Collodel:2022jly}, known as Q-balls~\cite{Coleman:1985ki}, which is an important feature of boson stars as mimickers. The concept of the boson star was first introduced by Kaup in the 1960s, who derived spherically symmetric solutions by coupling a complex scalar field to Einstein's gravity. Subsequently, Ruffini and Bonazzola conducted detailed investigations into similar models using a real scalar field~\cite{Kaup:1968zz,Ruffini:1969qy}. Since then, researchers have proposed various types of boson star solutions and achieved a series of significant advances, such as charged boson stars formed by coupling to electromagnetic fields \cite{Jetzer:1992tog}, Newtonian boson stars derived in the weak-field limit \cite{Silveira:1995dh}, and rotating boson stars with nonzero angular momentum \cite{Li:2019mlk}.

Polarization images play a crucial role in the study of black hole shadows. By comparing numerical polarization images with actual observational results, one can get deeper insight into the astrophysical characteristics and geometric structures of black hole accretion flows~\cite{Zhu:2022amy}. Taking the Schwarzschild black hole as a background, the EHT collaboration employed the approximate expression for null geodesics derived by Beloborodov~\cite{Beloborodov:2002mr} to obtain the polarization images of a black hole surrounded by hot gas~\cite{EventHorizonTelescope:2021btj}. This simplified model successfully reproduces the distribution of electric vector position angles and relative polarization intensity observed in the polarization image of M87*. Subsequently, this approach was further extended to Kerr black holes. Gelles et al. constructed a simplified model with an equatorial emission source, generated the corresponding polarization images, and analyzed the geometric effects of black hole spin on photon parallel transport~\cite{Gelles:2021kti}. These studies demonstrate that the polarization properties are primarily determined by the geometry of the magnetic field, while also being influenced by the intrinsic parameters of the black hole and the inclination of the observer. Beyond the Schwarzschild and Kerr scenarios, polarization images for various black hole models and horizonless ultra-compact objects have also been investigated extensively (see, for example, Refs.~\cite{Li:2025awg,Qin:2021xvx,Shi:2024bpm,Chen:2024jkm} and references therein). Notably, previous research has shown that for horizonless ultra-compact objects, such as traversable wormholes~\cite{Delijski:2022jjj} and naked Janis-Newman-Winicour singularities~\cite{Deliyski:2023gik}, the resulting polarization images display significant differences compared to those of black holes. Nevertheless, for boson stars under the thin accretion disk scenario, the properties of the polarization image remain uncertain.

Inspired by these studies, we will systematically investigate the polarization images of solitinic boson stars~\cite{Visinelli:2021uve}. Our primary objective is to reveal the influence of model parameters on the polarization images, as well as the correlation between polarization images and optical images of solitinic boson stars. This study facilitates the distinction between black holes and solitinic boson stars from the perspective of the polarization image.

\section{Solitonic Boson Star Model}
When a scalar field couples to gravity, a stable localized soliton structure can form by the self-interaction potential of the scalar field, which is the theoretical basis for the existence of solitonic boson stars~\cite{Visinelli:2021uve}. We consider the following solitonic boson star model, in which the self-gravitating (complex) scalar field is minimally coupled to gravity. The action is given by
\begin{equation}
	S=\int d^4x \sqrt{-\mathcal{g}} \left[\frac{R}{2 \kappa}-\nabla_b\Psi^{\star}\nabla^b\Psi-V(|\Psi|^2)\right],
\end{equation}
where $\mathcal{g}$ denotes the determinant of the metric, $R$ is the Ricci scalar, $\Psi^{\star}$ is the complex conjugate of the scalar field $\Psi$, and $V$ represents the self-interaction potential of the scalar field. Herefater, for convenience, we set the Einstein gravitational constant $\kappa = 8\pi G/c^4=1$. The sixtic potential in our solitonic boson star model is given by~\cite{Friedberg:1986tq}
\begin{equation}
	V(|\Psi|^2)= u^2 \Psi^2 \left(1-\frac{\Psi^2}{\alpha^2}\right)^2. \label{eq:vpsi}
\end{equation}
Here, $u$ denotes the mass of the scalar field, and $\alpha$ is a coupling parameter that controls the strength of the self-interaction. The property of the sixtic potential is that it vanishes at the false vacuum, where $\Psi = \alpha$. For large $\alpha$, the system behaves similarly to a mini boson star without self-interaction.

By varying the action with respect to the metric and scalar field independently, one obtains the  equations of motion for the model:
\begin{align}
	&R_{ab}-\frac{1}{2}Rg_{ab} =  T_{ab},\label{eq:efe1}\\
	&g^{ab}\nabla_b\nabla_c\Psi = \Psi \frac{dV}{d|\Psi|^2}, \label{eq:efe2}
\end{align}
where the energy-momentum tensor is given by
\begin{equation}
	T_{ab} = \nabla_a\Psi^{\star}\nabla_b\Psi + \nabla_b\Psi^{\star}\nabla_a\Psi - g_{ab} \left( \nabla_c\Psi^{\star}\nabla^c\Psi + V \right).
\end{equation}
We consider the scalar field with the following configuration:
\begin{equation}
	\Psi(r,t) = \psi(r) e^{i \omega t}. \label{eq:psi}
\end{equation}
To simplify the model, we neglect the rotation of the solitonic boson star. Therefore, the model is described by a static, spherically symmetric spacetime background, where the line element can be assumed as
\begin{equation}
	ds^{2} = -f(r)dt^{2} + g(r)^{-1}dr^{2} + r^{2}(d\theta^2 + \sin^2\theta\, d\varphi^2). \label{eq:metric}
\end{equation}
The two metric functions $f(r)$ and $g(r)$ can be given by solving the equations of motion. Substituting Eqs.~(\ref{eq:vpsi}), (\ref{eq:psi}), and (\ref{eq:metric}) into Eqs.~(\ref{eq:efe1}) and (\ref{eq:efe2}), the equations of motion reduce to the following system of differential equations:
\begin{align}
	\frac{d f}{dr} =&  r \left[ \frac{\omega^2 \psi^2}{g} + f \left( \frac{d \psi}{dr} \right)^2 - \frac{f}{g} u^2 \psi^2 \left( 1 - \frac{\psi^2}{\alpha^2} \right)^2 \right]\label{eq:me1}\\\nonumber&+\frac{f(1 - g)}{g\, r},\\
	\frac{d g}{dr} = &\frac{1 - g}{r} - \frac{ r \left[ f u^2 \psi^2 \left( 1 - \frac{\psi^2}{\alpha^2} \right)^2 + \psi^2 \omega^2 \right]}{f}\\\nonumber&- g \,r  \left( \frac{d \psi}{dr} \right)^2, \\
	\frac{d^2 \psi}{dr^2} =&- \frac{\psi \left[ u^2 \left( -\frac{3 \psi^4}{\alpha^4} + \frac{4 \psi^2}{\alpha^2} - 1 \right) + \frac{\omega^2}{f} \right]}{g}\label{eq:me3} \\\nonumber&-\frac{1}{2} \left( \frac{1}{f}\frac{df}{dr} + \frac{1}{g} \frac{dg}{dr} + \frac{4}{r} \right) \frac{d \psi}{dr}.
\end{align}
Due to the stiffness and complexity of Eqs.~(\ref{eq:me1})--(\ref{eq:me3}), obtaining analytical solutions is generally intractable. We therefore employ the shooting method for numerical integration. In this case, we need to first determine the boundary conditions for the functions $f(r)$, $g(r)$, and $\psi(r)$.

The equations of motion remain invariant under the following transformations
\begin{equation}
	r\to\mu\, r, \quad f\to\mu^2 f, \quad g\to g, \quad V\to\mu^{-2} V.
\end{equation}
Thus, without loss of generality, we can set $\mu = 1$ through appropriate scaling. At spatial infinity, the solution must asymptotically approach the flatness condition of the Schwarzschild spacetime, which yields
\begin{equation}
	\begin{aligned}
	&f(r\to\infty)= 1-\frac{2M}{r}|_{r\to\infty}\sim 1,\\ &g(r\to\infty)= 1-\frac{2M}{r}|_{r\to\infty}\sim 1, \\
&\psi(r\to\infty)=0.
	\end{aligned}
\end{equation}
Here, $M$ denotes the ADM mass of the solitonic boson star. Near the stellar center, i.e., $r \to 0$, the solution should remain regular.  Expanding each function as a polynomial series in $r$ and solving the resulting equations order by order, we can obatin the boundary conditions of the system at $r\to 0$, that is
\begin{equation}
	f(r \rightarrow 0) \sim f_0, \quad g(r \rightarrow 0) \sim 1, \quad \psi(r \rightarrow 0) \sim \psi_0,
\end{equation}
where $f_0$ and $\phi_0$ are two parameters related to the model. Since  $\phi_0$ governs the initial value of the scalar field $\Psi(r,t)$, we hereafter refer to it as the initial scalar field for convenience. Moreover, the equations of motion are independent of the coordinate time $t$, so we are free to reparametrize $t$, $\omega$, and $f(r)$ by a constant factor, which allows us to fix the parameter $f_0 = 1$. With these preliminaries established, we can investigate the polarization image of the solitonic boson star with a thin accretion disk. It is important to note that the direct application of the numerical metric to image boson stars located beneath the thin disk presents certain inconveniences. Consequently, we employ a fitting metric to address this issue \cite{He:2025qmq}.

\section{Linearly Polarized Light}

In this section, we investigate the linearly polarized light in the solitonic boson star model to advance our understanding of its radiation properties. We particularly focus on its interactions with the surrounding thin accretion disk.

We consider an optically and geometrically disk is situated within the equatorial plane, where the plasma contained within the disk follows circular orbits along timelike geodesics. Hence, the motion of particles in the accretion disk satisfies the condition $g_{\mu\nu}\dot{x}^\mu\dot{x}^\nu=-m^2$, where $m = 1$ denotes the  mass of the particle and $\dot{x}^\mu$ the derivative of spacetime coordinates with respect to the affine parameter $\tau$. In addition, the four-velocity of accretion flow is
\begin{align}\label{Fvelocity}
 u^\mu=\left(-\sqrt{\frac{2 g_{tt}^2}{r \partial_r g_{tt}-2 g_{tt}}}, 0, 0, \sqrt{\frac{r^3 \partial_r g_{tt}}{2 g_{tt}- r \partial_r g_{tt}}}\right).
\end{align}

Since a light ray may interact multiple times with the accretion disk before reaching the observer, the total observed intensity should include contributions from all possible interaction paths. By neglecting the reflection effect, the total observed light intensity can be expressed as
\begin{align} \label{II1}
\mathcal{J}_{\nu_{o}}=\sum^{N_m}_{n=1}g^3_n \mathcal{J}_n,
\end{align}
where $\nu_{o}$ is the observed frequency, and $N_m$ is the maximum number of intersections. Then, the redshift factor $g_n$ is
\begin{align} \label{redshift}
g_n=-\frac{1}{u^\mu g^{\mu\nu} p_\mu},
\end{align}
and $p_\mu$ is the four-momentum of photons.
For the emission profile $\mathcal{J}_n$, its functional form is adopted as~\cite{Gralla:2020srx}
\begin{align} \label{EMISS}
\mathcal{J}_n=\frac{e^{-\frac{1}{2}}[\gamma+\text{arcsinh}(\frac{r-\upsilon}{\sigma})]^2}{\sqrt{(r-\upsilon)^2+\sigma^2}}.
\end{align}
Here, $\gamma$ denotes the growth rate of the intensity, $\upsilon$ is the location of the intensity peak, and $\sigma$ governs the width of the profile. These three parameters ($\gamma$, $\upsilon$, $\sigma$ ) collectively characterize the thin accretion disk model, and their selection was originally designed to align with the results of GRMHD simulations of Kerr black holes.

Due to the placement of the accretion disk on the equatorial plane and its geometrically thin structure, it can be regarded as plasma moving along an equatorial timelike geodesic. Therefore, the emission of the polarized light originates from the synchrotron radiation emitted by electrons within the plasma. For an observer co-moving with the plasma, the polarization direction of the emitted light is perpendicular to both the local magnetic field $\vec{B}$ and the photon's three-momentum $\vec{k}$, so the the spatial component of the photon's polarization vector is calculated by~\cite{EventHorizonTelescope:2021btj,Gelles:2021kti,Delijski:2022jjj}
\begin{equation}
	\vec{f} = \frac{\vec{k} \times \vec{B}}{|\vec{k}|}. \label{eq:pv1}
\end{equation}
The generally covariant form of the photon's polarization vector is given by
\begin{equation}
	{f}^\mu \propto \xi^{\mu \nu \lambda \eta} \mu_\nu k_\lambda B_\eta, \label{eq:pv2}
\end{equation}
where the variables $\mu_\nu$, $k_\lambda$, and $B_\eta$ represent the four-velocity of the plasma, the four-momentum of the photon, and the magnetic field, respectively.
Furthermore, once the direction of the polarization vector is determined, it can be normalized to fulfill the conditions of orthogonality, that is
\begin{equation}
f^\mu f_\mu=1. \label{eqpv31}
\end{equation}
Within the framework of the emitter, the intensities of the linearly polarized light and the total light at the emission point can be described by the emission functions $\mathcal{J}_P$ and $\mathcal{J}_I$, respectively. According to Eq.~(\ref{II1}), we have
\begin{align} \label{II111}
	\mathcal{J}_I=\sum^{N_m}_{n=1} \mathcal{J}_n,
\end{align}
For simplicity, the emission intensity is assumed to be dependent solely on the emission point, and independent of both the photon frequency and the magnetic field. Hence, the emission intensity is only a function of the radial position $r$, and we can express the intensity $\mathcal{J}_P$ of the linearly polarized light in the form of $\mathcal{J}_I$, i.e.,
\begin{equation}
\mathcal{J}_P=C_0 \mathcal{J}_I(r). \label{eq:pv32}
\end{equation}
Here, $C_0 $ is a variable that describes the proportion of the linearly polarized light at the emission position in the total light intensity, which satisfies
\begin{equation}
0\leq C_0\leq1. \label{del}
\end{equation}
For the case of $C_0=1$, the emitted light is completely linearly polarized.

The observed polarization properties are determined by the parallel transport of the polarization vector $f^\mu$ along the null geodesic to the observer~\cite{Gelles:2021kti}. Therefore, the polarization vector $f^\mu$ satisfies \begin{equation}
k^\nu \nabla_\nu f^\mu=0. \label{PT}
\end{equation}
By introducing the affine parameter $\varsigma$, it can be specifically expressed as
\begin{equation}
\frac{d}{d \varsigma}f^\mu+\Gamma^\mu_\nu k^\nu f^\lambda=0. \label{PT2}
\end{equation}
The observed intensity $\mathcal{J}_{P\nu_o}$ of the linearly polarized light and the observed total intensity $\mathcal{J}_{\nu_{o}}$ can be expressed as
\begin{equation}
\mathcal{J}_{P\nu_o}=g^3 \mathcal{J}_P=C_0\,g^3 \mathcal{J}_I(r), \quad \mathcal{J}_{\nu_{o}}=g^3 \mathcal{J}_I, \label{PT3}
\end{equation}
where $g$ represents the overall redshift factor.

To obtain the polarization image of the solitonic boson star, it is also necessary to select an appropriate observer and set up a corresponding imaging screen. A natural choice is to consider an observer with vanishing axial angular momentum, commonly referred to as a zero angular momentum observer (ZAMO). Specifically, the ZAMO is positioned at coordinates $(t_o, r_o, \theta_o, \varphi_o)$, and a locally orthogonal normalized frame exists in the neighborhood of the observer, which can be written as
\begin{align}
	&e_{(t)} =\left(\sqrt{-\frac{1}{g_{tt}}}, 0, 0, 0\right),\quad e_{(r)} =\left(-\sqrt{\frac{1}{g_{rr}}}, 0, 0, 0\right),\label{eq:tetrad1}\\
	&e_{(\theta)} =\left(0, \sqrt{\frac{1}{g_{\theta \theta}}}, 0, 0\right),\quad e_{(\varphi)} =\left(-\sqrt{\frac{1}{g_{\varphi \varphi}}}, 0, 0, 0\right). \label{eq:tetrad2}
\end{align}
Within the framework of the ZAMO, the four-momentum of the photon is expressed as $ p_{(\mu)} = p_{\nu} e^{\nu}_{(\mu)} $, where $ p_{(\mu)} $  and $ p_{\nu} $ denote the four-momentum in the ZAMO  frame and the Boyer-Lindquist coordinate system. At the observer's location, it is essential to define a celestial sphere coordinate system $(X , Y)$ and then transform the celestial sphere coordinates into the corresponding points on the imaging plane $(\tilde{\alpha}, \tilde{\beta})$ within the Cartesian coordinate system. The relationship between the four-momentum $p_{(\mu)}$ of the photon and the celestial sphere coordinates $(X, Y)$ is established by the following equations~\cite{Hu:2020usx}:
\begin{align} \label{TM3}
\cos X=\frac{p^{(r)}}{p^{(t)}}, \qquad \tan Y= \frac{p^{(\varphi)}}{p^{(\theta)}}.
\end{align}
On the imaging plane $(\tilde{\alpha}, \tilde{\beta})$, the transform between the Cartesian coordinates and the sphere celestial coordinates satisfies
\begin{align}\label{TM4}
	\tilde{\alpha} = -2\tan\frac{X}{2}\sin Y,\\
	\tilde{\beta} = -2\tan\frac{X}{2}\cos Y.
\end{align}

In this work, we focus on two observable quantities in the polarization image, i.e., the direction of the polarization vector on the observer's celestial sphere and the polarization intensity. On the imaging plane, the projection of the polarization vector satisfies
\begin{align}\label{PV1}
f^{(\tilde{\alpha})} =f^u \cdot e_{\tilde{\alpha}}=-f^u\cdot e_\varphi,\quad f^{(\tilde{\beta})} =f^u \cdot e_{\tilde{\beta}}=-f^u\cdot e_\theta.
\end{align}
In fact, both $\vec{f}$ and $-\vec{f}$ describe the same linearly polarized light. Thus, to standardize the direction of the polarization vector, we can set $f^{(\tilde{\beta})}>0$ and $\Phi_{\mathrm{EVPA}} \in (0, \pi)$, where $\Phi_{\mathrm{EVPA}}$ is the electric vector position angle (EVPA). The total intensity of the linearly polarized light detected by the observer is obtained by summing the contributions of the linearly polarized emission from all points on the equatorial plane. According to the definitions of the Stokes parameters $\mathcal{Q}$ and $\mathcal{U}$~(see Ref.~\cite{Huang:2024bar}), which obey the principle of linear superposition, the total intensity of the linearly polarized light detected by the observer can be expressed as
\begin{align}\label{PV2}
\mathcal{Q}_{\mathrm{all}} &= \sum_{n=1}^{N_m} \mathcal{Q}_{n} = \sum_{n=1}^{N_m} g_{n}^{3} \mathcal{J}_{P_n} \left[ (f_n^{(\tilde{\alpha})})^2 - (f_n^{(\tilde{\beta})})^2 \right], \\
\mathcal{U}_{\mathrm{all}} &= \sum_{n=1}^{N_m} \mathcal{U}_{n} = \sum_{n=1}^{N_m} g_{n}^{3} \mathcal{J}_{P_n} \left( 2 f_n^{(\tilde{\alpha})} f_n^{(\tilde{\beta})} \right).
\end{align}
Hence, the observed total polarization intensity and the EVPA are given by
\begin{align}\label{PV3}
\mathcal{J}_{P\nu_o} = \sqrt{\mathcal{Q}_\mathrm{all}^2 + \mathcal{U}_\mathrm{all}^2},	
\end{align}
and
\begin{align}\label{PV4}
\Phi_{\mathrm{EVPA}} &= \frac{1}{2} \arctan \frac{\mathcal{U}_{\mathrm{all}}}{\mathcal{Q}_{\mathrm{all}}}.
\end{align}
Up to this point, the polarization image of the solitonic boson star can be numerically simulated on the imaging plane. The impact of the initial scalar field $\phi_0$ and the coupling parameter $\alpha$ on the polarization image will be discussed in the next section.

\section{Numerical Results of the Polarization Image}
Given that the observations of M87* suggest that the magnetic field configuration $\vec{B} = (0.87, 0.5, 0)$ can accurately account for its polarization features, we employ these same parameters to simulate the polarization characteristics of solitonic boson stars. In our prior study~\cite{He:2025qmq}, we have performed a comprehensive analysis of the optical images of the same solitonic boson stars with a same thin accretion disk, systematically elucidating the evolution of the optical image in response to variations in the initial scalar field $\phi_0$ and the coupling parameter $\alpha$. Therefore, while investigating the influence of model parameters on the polarization image, we can also get deeper insights into the observational characteristics of solitonic boson stars by comparing these two types of images.

\begin{figure}[h]
\centering
\includegraphics[width=13cm]{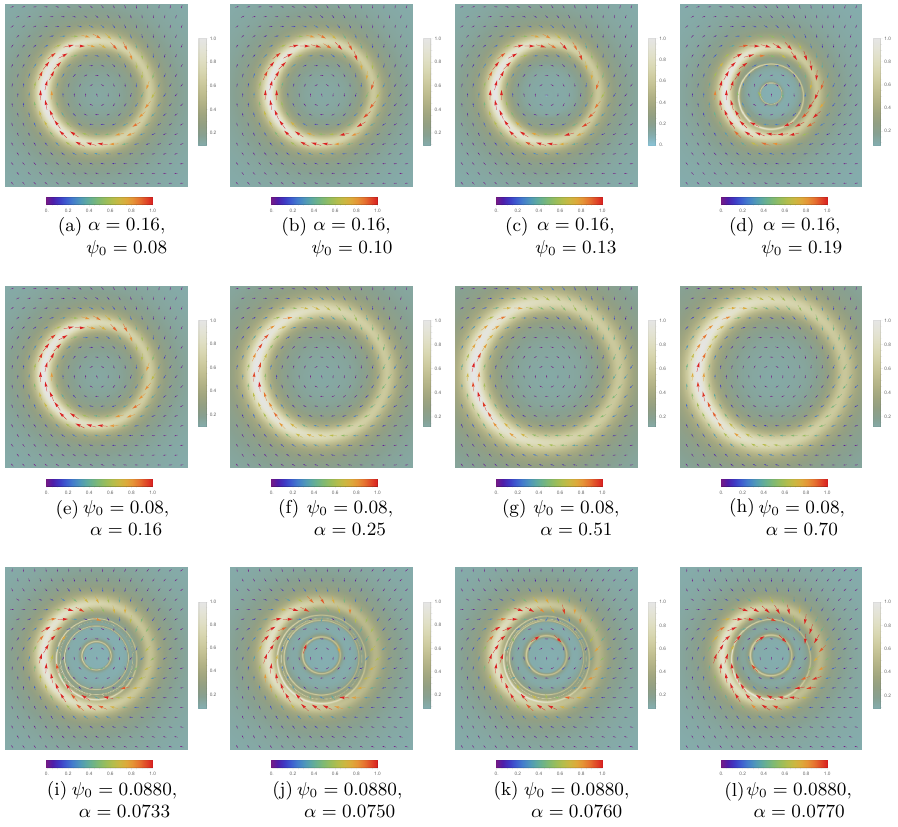}
\caption{Polarization images of solitonic boson stars. The colored arrows represent the linearly polarized vectors, with their length and color depth proportional to the polarization intensity $|\mathcal{J}_{P\nu_o}|$. The orientation reflects the electric vector position angle $\Phi_{\mathrm{EVPA}}$. The observation inclination angle  is set to $\theta_o = 17^\circ$.}
	\label{fig1}
\end{figure}

\begin{figure}[h]
	\centering
	\includegraphics[width=13cm]{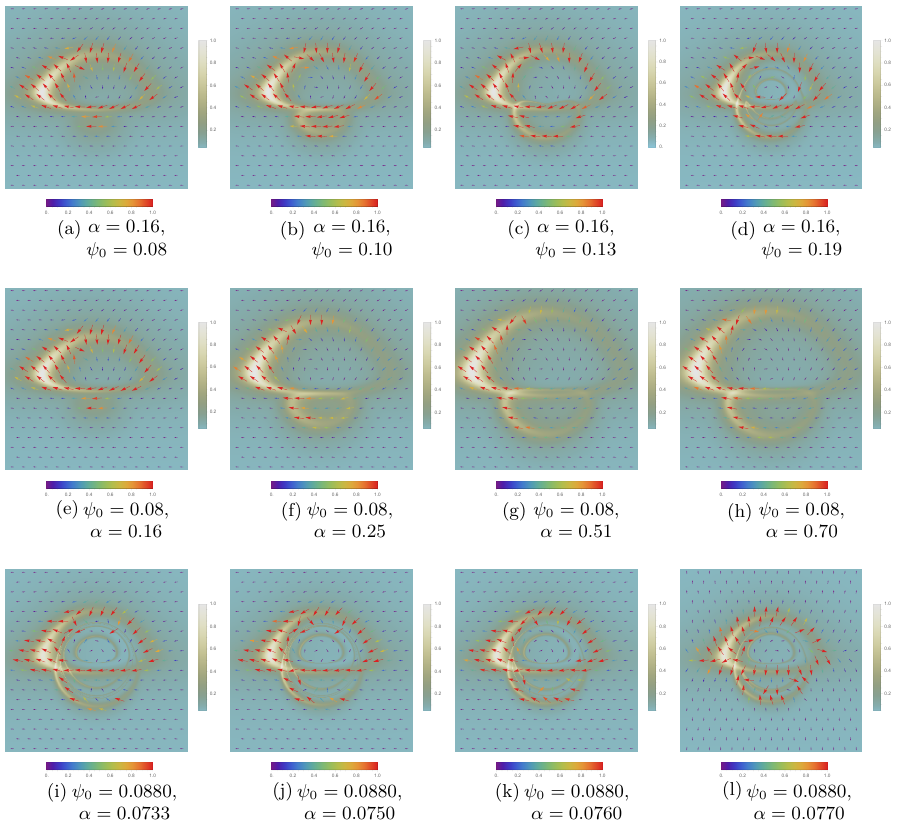}
	\caption{Polarization images of solitonic boson stars. The colored arrows represent the linearly polarized vectors, with their length and color depth proportional to the polarization intensity $|\mathcal{J}_{P\nu_o}|$. The orientation reflects the electric vector position angle $\Phi_{\mathrm{EVPA}}$. The observation inclination angle  is set to $\theta_o = 75^\circ$.}
	\label{fig2}
\end{figure}

In Figs.~\ref{fig1} and~\ref{fig2}, we plot the polarization images of solitonic boson stars for different values of the initial scalar field $\psi_0$ and the coupling parameter $\alpha$. Note that, for the optical image of a compact star, light rays traversing the accretion disk different numbers of times produce different images. Specifically, the primary image (direct image) is associated with the case of $n = 1$, the secondary image (lens image) corresponds to $n = 2$, and three or more intersections represent the higher-order image (photon ring). In previous literature, the definition of the photon ring varies. For example, in Refs.~\cite{Gralla:2019drh,Hou:2022gge}, all images located outside the direct image are collectively referred to as the photon ring. However, in Refs.~\cite{Gralla:2019xty,He:2024amh}, the term of the photon ring is used to describe the images that lie beyond both the primary and secondary images. In this work, we adopt the latter definition. In Fig.~\ref{fig1}, the observation inclination angle is set to $\theta_o = 17^\circ$, which is consistent with the viewing angle of M87* inferred by the EHT. As a comparison, we plot Fig.~\ref{fig2} presenting the corresponding polarization images at a larger observation inclination angle ($\theta_o = 75^\circ$).

Now, let us analyze the information provided by these polarization images. As shown in Fig.~\ref{fig1}, compared with the optical images given in Ref.~\cite{He:2025qmq}, we can observe that, for all values of the parameters $\phi_0$ and $\alpha$, the polarization intensity $|\mathcal{J}_{P\nu_o}|$ at the location of the direct image is always the most prominent. There is a positive correlation between the polarization intensity of the polarization image and the brightness of the optical image. This characteristic is independent of the observation inclination angle (see Fig.~\ref{fig2} with $\theta_o = 75^\circ$).

From the images and parameter values in the first row (with fixed $\alpha=0.16$), it can be observed that as the initial scalar field $\phi_0$ increases, the lensed image and photon ring emerge, accompanied by enhanced polarization at their corresponding positions. However, the overall polarization intensity at these locations remains relatively weak compared to that of the direct image. When the coupling strength is small (see the second row, corresponding to large values of the coupling parameter $\alpha$), as the coupling further decreases (see from left to right), the polarization intensity weakens, and the polarization intensity distribution primarily concentrates on the left side of the direct image. However, when the coupling strength becomes sufficiently large (see the bottom row, associated with small values of $\alpha$), the lensed image and photon ring appear. In this case, as $\alpha$ increases (see from left to right), the polarization intensity distribution begins to spread more uniformly across the direct image. This indicates that the influence of $\alpha$ on polarization intensity distribution is not monotonic. Moreover, the directions of all the arrows in Fig.~\ref{fig2} (which reflect the variations in $\Phi_{\mathrm{EVPA}}$) show no significant changes with variations in either $\alpha$ or $\phi_0$.

In Fig.~\ref{fig2}, we illustrate a scenario with a larger inclination angle $\theta_o$. The external ``cap-shaped'' bright ring corresponds to the direct image, whereas the internal  ``D-shaped'' ring represents the lensed image. This observation inclination angle further confirms a positive correlation between the polarization intensity distribution in the polarization images and the brightness in the optical images. As for the polarization intensity distribution, the influence of the coupling strength $\alpha$ is also consistent with the results observed at the small observation inclination angle. A particularly noteworthy observation is that the arrow directions in Fig.~\ref{fig2} (l) have undergone a significant change in the central region of the direct image compared to other images, which is a phenomenon not evident in Fig.~\ref{fig1} (l). This contrast clearly implies that the observation inclination angle substantially affects the visibility of the variations in $\Phi_{\mathrm{EVPA}}$.

Finally, we examine the differences between the polarization images of solitonic boson stars and those of black holes. In the case of black holes, the presence of the event horizon prevents polarization vectors from appearing within the interior region~\cite{Zhang:2022klr}. In contrast, solitonic boson stars exhibit no event horizon, allowing polarization vectors to appear inside the star (see Figs.~\ref{fig1} and \ref{fig2}). The resulting observational signatures  provide a unambiguous diagnostic feature for identifying solitonic boson stars. Furthermore, at large observation inclination angles, specific combinations of the coupling parameter and the initial scalar field (for example $\alpha=0.0770$ and $\phi_0=0.0880$) can produce anomalous  $\Phi_{\mathrm{EVPA}}$ in localized regions of the direct image. These distinctive polarization features are not typically observed in black hole systems and they may also serve as a distinguishing criterion between black holes and solitonic boson stars.

\section{Conclusion}
This study systematically analyzes the polarization characteristics of solitonic boson stars in a thin accretion disk scenario through numerical simulations. The polarization intensity distribution is significantly positively correlated with the brightness distribution of the optical image. The strongest polarization consistently appears at the location corresponding to the direct image, which is independent of the observation inclination angle. The effect of the coupling strength parameter $\alpha$ on the polarization intensity distribution is non-monotonic: under weak coupling (large $\alpha$), the polarization intensity decreases and shifts leftward as $\alpha$ increases; under strong coupling (small $\alpha$), the polarization intensity increases and becomes uniformly distributed across the entire direct image as $\alpha$ increases.

Due to the absence of the event horizon in solitonic boson stars, polarization vectors can penetrate the interior of the star (see Figs.~\ref{fig1} and \ref{fig2}), whereas no polarization signals are observed within the event horizon of black holes. Moreover, at high observation inclination angles, certain parameter combinations  (for example $\alpha=0.0770$ and $\phi_0=0.0880$) can induce anomalous  $\Phi_{\mathrm{EVPA}}$ in localized regions of the direct image [see Fig.~\ref{fig2} (l)], which do not occur in black holes. These polarization characteristics of solitonic boson stars reveal their fundamental differences from black holes and lay the theoretical foundation for next-generation EHT observations to distinguish solitonic boson stars from black holes.

In the future, we can explore the polarization response of rotating (solitonic) boson stars, higher-order potential functions, or dynamic scalar fields to test the universality of the results. By combining data from X-ray polarimeters (such as IXPE) and next-generation EHT, one can verify the applicability of the polarizationrightness correlation across different spectral bands.

\vspace{10pt}

\noindent {\bf Acknowledgments}

\noindent
This work is supported by the National Natural Science Foundation of China (Grant No. 12375043), the Natural Science Foundation of Chongqing (CSTB2023NSCQ-MSX0594), and the China Postdoctoral Science Foundation (Grant No. 2024M753825).



\end{document}